\documentstyle[preprint,aps,psfig]{revtex}
\begin{document}

\tighten

\preprint{\vbox{
\hbox{CWRU-P2-1996}
}}

\title{SU(5) Monopoles and the Dual Standard Model}

\author{Hong Liu\footnote{hxl20@po.cwru.edu} and 
Tanmay Vachaspati\footnote{tanmay@theory4.phys.cwru.edu}}
\address{
Physics Department\\
Case Western Reserve University\\
Cleveland OH 44106-7079.
}

\date{\today}

\maketitle

\begin{abstract}
We find the spectrum of magnetic monopoles 
produced in the symmetry breaking 
$SU(5) \rightarrow [SU(3)\times SU(2)\times U(1)']/Z_6$ 
by constructing classical bound states of the fundamental 
monopoles. The spectrum of monopoles is found to correspond 
to the spectrum of one family of 
standard model fermions and hence, is a
starting point for constructing the dual standard model. 
At this level, 
however, there is an extra monopole state - the ``diquark'' 
monopole - with no corresponding standard model fermion. 
If the $SU(3)$ factor now breaks down to $Z_3$, the monopoles 
with non-trivial $SU(3)$ charge get confined by strings in 
$SU(3)$ singlets. Another outcome of this symmetry breaking 
is that the diquark monopole becomes unstable (metastable) to 
fragmentation into fundamental monopoles and the one-one 
correspondence with the standard model fermions is restored.
We discuss the fate of the monopoles if the 
$[SU(2)\times U(1)']/Z_2$ factor breaks down to $U(1)_Q$ by a 
Higgs mechanism as in the electroweak model. Here we find that 
monopoles that are misaligned with the vacuum get connected by 
strings even though the electroweak symmetry breaking does not 
admit topological strings. We discuss the lowest order quantum
corrections to the mass spectrum of monopoles.

\end{abstract}

\pacs{}


\section{Introduction}

The idea that particles and solitons are two descriptions of
the same entities has an irresistible appeal: it promises to
unify seemingly distant aspects of field theories and also
to resolve various mysterious relations that might hold in a
description of particles or solitons alone. Research over the
past several decades \cite{skyrme,sinegordon} and 
(more intensely) over the last few years \cite{intril} has 
supported the particle-soliton duality and it appears
that there is some truth in this idea. At the same time, the
research has focussed on hypothetical field theory models where
exact results can be derived. In this paper, we wish to follow
up on the ideas presented in \cite{tv} and consider how duality
might be relevant to the world we know.

It is well known that a
theory with spontaneously broken symmetry can lead to
magnetic monopole solutions \cite{thap} provided the
vacuum manifold of the theory contains incontractible
two spheres. Specifically, if $G \longrightarrow K$
denotes the symmetry breaking pattern where the initial
symmetry group $G$ is assumed to be simply connected, the
monopoles are classified by the incontractible paths in $K$.
This yields the topological classification of magnetic
monopoles which has been repeatedly used over the last two
decades to classify magnetic monopoles in field
theories \cite{godolive}.
In spite of the success and widespread use of the topological
classification of magnetic monopoles,
there are reasons to believe that a
classification of monopoles exists \cite{goddardetal} that
is finer than the topological classification.
In the finer classification the monopoles form
representations of a certain dual symmetry group $K^v$
which has been explicitly worked out in a number of cases 
\cite{goddardetal,danieletal}.
A monopole that appears to be the sole member in a given topological
class is actually an irreducible multiplet of $K^v$ and the
field theory of monopoles has a $K^v$ gauge symmetry.

These considerations have immediate application in the study of 
$SU(5)$ monopoles where the symmetry breaking one considers is:
\begin{eqnarray*}
G = SU(5) \longrightarrow 
H =  [SU(3)\times SU(2)\times U(1)']/Z_6 \ .
\end{eqnarray*}
The spectrum of {\it stable} monopoles resulting from this symmetry 
breaking has been found \cite{tv} to be in correspondence 
with the fermions of a single family of the standard model and the
dual group $H^v$ is locally isomorphic to $H$ \cite{danieletal}.
Hence the stable monopoles of the purely bosonic $SU(5)$ model may 
just be a different description of the fermionic particles that are 
the constituents of the world we know.
Our hope is that the present direction of research might
eventually lead to a ``dual standard model'' which would be an
alternate description of particle physics 
and would be immediately relevant to the strongly coupled 
QCD sector. A comparison
of some of the features of the standard model versus the dual
standard model is made in Table I.

It is worthwhile clarifying our use of the word ``dual''.
What we have in mind is that the monopoles of the $SU(5)$ model are
simply a different description of the {\it fermions} in the standard
model. In other words, the standard model is an effective 
theory of fields that create and annihilate $SU(5)$ monopoles. 
This is analogous to the sine-Gordon and Thirring model 
equivalence.

\vfill

\begin{center}
TABLE I.
\small
A comparison of the standard and the dual standard model.
($g_3$, $g_2$ and $g_1$ refer to the coupling constants of
$SU(3)$, $SU(2)$ and $U(1)$ respectively and the tildes denote
the same quantities in the dual representation.)
Both descriptions have a strong coupling problem - the
standard model has quark confinement, the dual standard
model has electroweak magnetic screening as described in
Sec. VI.
\end{center}
\normalsize
\begin{center}
\begin{tabular*}{10.0cm}{|c|@{\extracolsep{\fill}}c|}
\hline
&\\[-0.1cm]
                     {Standard Model} &
                     {Dual Standard Model} \\[0.2cm]
\tableline
&\\
$SU(3)$&${\tilde{SU}} (3)$                   \\
$g_3 > 1$ & ${\tilde g}_3 < 1$        \\
Quark confinement & String confinement       \\
&\\
\hline
&\\
$SU(2)\times U(1)$&${\tilde {SU}}(2)\times {\tilde U}(1)$        \\
$g_2 \ , g_1 < 1$
& $\tilde{g}_2 \ , {\tilde g}_1 > 1$              \\
Higgs screening & Electroweak magnetic screening \, \\
&\\
\hline
\end{tabular*}
\end{center}

\vspace{0.3cm}

We begin in Sec. II by constructing the entire spectrum of stable
magnetic monopoles in an $SU(5)$ model. Here we find the existence
of monopoles that are in correspondence with a single family
of fermions of the standard model; in addition we find an extra
monopole that we call the ``diquark'' monopole since it is a bound state
of two (quark like) fundamental monopoles with twice the color
charge. The classical mass spectrum of the stable monopoles is also
determined in the weak coupling limit.
In Sec. III we qualitatively consider quantum corrections to the mass
spectrum of the monopoles and provide some partial results. 

The correspondence between monopoles and standard model fermions
requires that colored monopoles be confined by strings just like
colored quarks are confined by QCD strings \cite{nambu1}. 
In Sec. IV we discuss
how the confinement of ``colored'' monopoles is achieved
by further breaking the ``dual color'' symmetry. At this stage,
as described in Sec. V, we find that the diquark monopole becomes
unstable (metastable) to fragmentation into fundamental monopoles
and the correspondence of stable monopoles and standard model
fermions becomes one to one.

We discuss the effect of the ``dual electroweak'' symmetry breaking
on the $SU(5)$ monopoles in Sec. VI. In particular, we have tried to
resolve the issue of what happens to $SU(5)$ monopoles when
the dual electroweak symmetry group breaks down to the dual
electromagnetic group. Based on classical arguments we 
conclude that monopoles whose magnetic 
charge is not purely electromagnetic get connected by strings
even though the vacuum manifold for the electroweak symmetry
breaking has trivial first homotopy. We discuss the limitation
of the classical calculations we have used in the context of 
constructing the dual standard model.

In Sec. VII we discuss the issue of spin of the monopoles.
We propose a simple scheme based on work done in the
70's \cite{jr,goldhaber} to convert the monopoles into
fermions. This scheme generates extra light dyonic states
which can be eliminated by introducing a $\theta$ term in the
action.

We summarize our findings in Sec. VIII and discuss future
directions.

\section{Spectrum of Stable Monopoles}

Consider the symmetry breaking
\begin{equation}
G\equiv SU(5) \longrightarrow [SU(3)\times SU(2)\times U(1)']/Z_6
\equiv H_{SM}
\label{symbreaking}
\end{equation}
The subscript $SM$ on $H$ denotes that this group is also
the symmetry group of the Standard Model. The $Z_6$ factor
is actually the direct product $Z_3 \times Z_2$ where
$Z_3$ is the center of $SU(3)$ and $Z_2$ is the center of
$SU(2)$. The symmetry breaking
can be realized when a Higgs field ($\Phi$) in the adjoint
representation of $SU(5)$ acquires a vacuum expectation value
(VEV). The Higgs potential for $\Phi$ can be written as
\begin{equation}
V(\Phi ) = -m_1 ^2 ({\rm Tr}\Phi^2) + a ({\rm Tr} \Phi^2 )^2
            + b {\rm Tr}(\Phi^4) \
\label{higgspotential}
\end{equation}
and the expectation value of $\Phi$ can be taken to be
\begin{equation}
<\Phi > = v_1 {\rm diag}(2,2,2,-3,-3)\
\label{vevofphi}
\end{equation}
where, $v_1=m_1/\sqrt{60a+14b}$.
After symmetry breaking, the scalar field $\Phi$ can
be decomposed into pieces that transform in the
$({\bf 8},{\bf 1})$, $({\bf 1},{\bf 3})$ and
$({\bf 1},{\bf 1})$ representations of $SU(3)\times SU(2)$
(for a review see \cite{chengli14}).
The masses of these scalar fields are denoted by
$\mu_8 = \sqrt{20 b}\ v_1$, $\mu_3 = 2\mu_8$ and $\mu_0 = 2 m_1$
respectively. We shall be interested in the case $\mu_0 << \mu_8$.

The topological classification of the monopoles in this
symmetry breaking is based on the second homotopy group
\begin{equation}
\pi_2 ( SU(5)/[SU(3)\times SU(2)\times U(1)']/Z_6 )
\cong \pi_1 ([SU(3)\times SU(2)\times U(1)']/Z_6 ) \ .
\label{pi2}
\end{equation}
Hence the topologically inequivalent monopoles in this case
can be found by considering the incontractible closed
paths in $H$. The set of incontractible paths is not empty since
$U(1)'$ is not simply connected. Yet the path that lies entirely
in $U(1)'$ does not yield the fundamental monopole - the monopole
that has the smallest non-vanishing $U(1)'$ charge. This is
because the center of $SU(3)\times SU(2)$ is given by the
group elements,
$$
\{ {\bf 1}_3, e^{i2\pi /3} {\bf 1}_3 , e^{i4\pi /3} {\bf 1}_3 \}
\oplus \{ {\bf 1}_2, e^{i\pi} {\bf 1}_2 \}
$$
which also belong to $U(1)'$. (${\bf 1}_n$ denotes the $n\times n$
unit matrix.) Hence a path that goes around the $U(1)'$ one sixth
of the way can be closed off by elements of the $SU(3)$ and $SU(2)$
groups. Specifically, an incontractible path yielding the
fundamental monopole can be written as:
\begin{equation}
g(s) = exp[ i M_1 s ]\ , \quad s \in [0,2\pi ]
\label{fundpath}
\end{equation}
where,
\begin{equation}
M_1 = Q_3 + Q_2 + Q_1 \ ,
\label{m1}
\end{equation}
with,
\begin{equation}
Q_3 = {\rm diag}
\left ( -{1\over 3}, -{1\over 3}, +{2\over 3}, 0,0 \right )
\label{q3}
\end{equation}
\begin{equation}
Q_2 = {\rm diag} \left ( 0,0,0, +{1\over 2},-{1\over 2} \right )
\label{q2}
\end{equation}
\begin{equation}
Q_1 = {\rm diag} \left ( +{1\over 3}, +{1\over 3},
         +{1\over 3}, -{1\over 2},-{1\over 2} \right ) \ .
\label{q1}
\end{equation}
We can explicitly obtain the charge on the fundamental monopole
from (\ref{m1}):
\begin{equation}
M_1 = {\rm diag} (0,0,1,0,-1) \ .
\label{q1exp}
\end{equation}

A path that traverses the whole $U(1)'$ is given by
$$
e^{i6 Q_1 s} \ , \quad s \in [0,2\pi ]
$$
and, in this sense, the fundamental monopole only
traverses one sixth of the full $U(1)'$ orbit. So the
fundamental monopole has $1/6$ of magnetic $U(1)'$ charge.
Similarly, a path that traverses the whole $U(1)$ subgroup
of $SU(2)$ generated by $Q_2$ is given by
$$
e^{i2 Q_2 s} \ , \quad s \in [0,2\pi ]
$$
and so the fundamental monopole has $1/2$ of magnetic
$SU(2)$ charge. By identical arguments we find that the 
fundamental monopole has $1/3$ of magnetic $SU(3)$ charge.

Note that as long as the group $G$ in (\ref{symbreaking})
is simply connected, the topological classification of
monopoles is given by $\pi_1 (H_{SM})$ and is
independent of $G$ itself. So the same spectrum of
monopoles will be produced for any simply connected
$G$ that breaks down to $H_{SM}$.

It has been conjectured for some time now \cite{goddardetal}
that the topological classification of magnetic monopoles
is not the entire story and a finer classification
of magnetic monopoles exists. In this finer
classification scheme, the magnetic monopoles form
a representation of a dual group. The
existence of multi-dimensional representations of the
fundamental monopole, for example, means that this
monopole should not be regarded as a single monopole
but as a member of a multiplet containing several
degrees of freedom. In the specific case of $SU(5)$
fundamental monopoles, this multiplet corresponds to
six different ways of writing the charge on the
monopole which correspond to the six different ways of
aligning the monopole charge in internal space.
This is seen by realizing that the diagonal $SU(3)$ generator
in (\ref{q3}) can be written in three different ways
by putting the $+2/3$ in one of three different
positions and the $SU(2)$ generator in (\ref{q2})
can be written in two different ways by putting
the $+1/2$ in one of two different positions.
We will denote the three different $SU(3)$ generators
by $T_c$ where $c$ can be {\it b}lue, {\it g}reen or {\it r}ed,
the two different $SU(2)$ generators by $\lambda_+$ and
$\lambda_-$ and the $U(1)$ generator by $Y$. The explicit
matrices are:
\begin{equation}
T_b = {\rm diag}
\left ( -{1\over 3}, -{1\over 3}, +{2\over 3}, 0,0 \right )
\label{tb}
\end{equation}
\begin{equation}
T_g = {\rm diag}
\left ( -{1\over 3}, +{2\over 3}, -{1\over 3}, 0,0 \right )
\label{tg}
\end{equation}
\begin{equation}
T_r = {\rm diag}
\left ( +{2\over 3}, -{1\over 3}, -{1\over 3}, 0,0 \right )
\label{tr}
\end{equation}
\begin{equation}
\lambda_+ = {\rm diag} \left ( 0,0,0, +{1\over 2},-{1\over 2} \right )
\label{lplus}
\end{equation}
\begin{equation}
\lambda_- = {\rm diag} \left ( 0,0,0, -{1\over 2},+{1\over 2} \right )
\label{lminus}
\end{equation}
\begin{equation}
Y = {\rm diag} \left ( +{1\over 3}, +{1\over 3},
         +{1\over 3}, -{1\over 2},-{1\over 2} \right ) \ .
\label{y}
\end{equation}
Note that
\begin{equation}
T_b+T_g+T_r = 0
\label{bgr}
\end{equation}
and
\begin{equation}
\lambda_+ +\lambda_- = 0 \ .
\label{pm}
\end{equation}

With this notation, the six different
fundamental monopole states can be labeled by their
$SU(3)$, $SU(2)$ and $U(1)'$ charges:
\begin{equation}
|c,s,1> \ , \quad c=b,g,r \ , \quad s=+,-
\label{fundmono}
\end{equation}
and the charge on the monopole is
\begin{equation}
Q_m = T_c + \lambda_s + Y \ .
\label{fmcharge}
\end{equation}

It should be remembered that all these six monopoles are
topologically equivalent. But we should still treat them
as being distinct because when we start combining them to
form higher winding monopoles, the stability of the resulting
monopole depends crucially on which two monopoles we consider.
In other words, the interaction energy of two monopoles - which
is a gauge invariant quantity - distinguishes between these
monopoles\footnote{One can also consider other ways of distinguishing
between the monopoles. For example, in the
event that the monopoles enter a region with magnetic
field, the trajectory that the monopoles follow will be
different for the different monopoles.}.

The existence of six different fundamental monopoles has
an important consequence that was first utilized by
Gardner and Harvey \cite{gh} in the context of $SU(5)$
monopoles. If one attempts to construct
charge two magnetic monopoles by combining two identical
charge one monopoles, one can show that the resulting
configuration is unstable due to the Coulombic repulsion
between the monopoles. (In the critical
Bogomolnyi case \cite{bogo}, the configuration is neutrally
stable.) However, one can still combine non-identical
fundamental monopoles and hope to obtain a stable charge
two monopole. Since we have six non-identical monopoles,
we have the possibility of getting $^6C_2 = 15$ different
charge two monopoles that are stable. But not all these
15 monopoles will be stable. To decide which of the 15
combinations will be stable, we follow \cite{gh} and
construct the interaction energies of two monopoles
whose charges are written as:
\begin{equation}
M_1 = n_1 Y + n_3 \lambda_i + n_8 T_a
\end{equation}
\begin{equation}
M_2 = n_1' Y + n_3 ' \lambda_{j} + n_8 ' T_{b}  
\end{equation}
where the $n_i$ and $n_i'$ are some integers. At separations
larger than the monopole core size, the
monopoles interact via gauge boson and scalar exchange
and the interaction energy is:
\begin{equation}
V(r) = {1 \over {4\alpha r}} [
       n_1 n_1' {\rm Tr}(Y^2) (1-e^{-\mu_0 r}) +
       n_3 n_3' {\rm Tr}(\lambda_i \lambda_j) (1-e^{-\mu_3 r}) +
       n_8 n_8' {\rm Tr}(T_a T_b) (1-e^{-\mu_8 r}) ] \
\label{intenergy}
\end{equation}
where $\alpha$ is the $SU(5)$ fine structure constant.
At separations less than the core size, it is assumed that
the monopoles interact like Bogomolnyi-Prasad-Sommerfield
(BPS) \cite{bogo} monopoles and hence the interaction
potential is flat. If the core size is very small compared
to the other length scales, we can approximate the true
interaction potential by (\ref{intenergy}) all the way down to $r=0$.
The monopole resulting from the combination of the two monopoles
will be classically stable provided $V(r)$ is in the shape of a
potential well at $r=0$. Therefore, since $V(r)$ goes to zero as $r$
goes to infinity, quantum stability can only be obtained if $V(r)$ is
negative at the origin\footnote{We are restricting ourselves to cases 
where there is classical stability and there is also a chance that the
monopole will be quantum mechanically stable. Metastable
states - states that can decay by quantum tunneling - are not being
considered here.}.
This reduces the problem to determining the monopoles for which
\begin{equation}
V(0) = {1 \over {4\alpha}} [
       n_1 n_1' {\rm Tr}(Y^2) \mu_0 +
       n_3 n_3' {\rm Tr}(\lambda_i \lambda_j) \mu_3 +
       n_8 n_8' {\rm Tr}(T_a T_b) \mu_8  ] \ < \ 0 \ .
\label{intenergyatzero}
\end{equation}

Let us start by constructing $n=2$ monopoles by combining
two $n=1$ monopoles. Then $n_i = 1= n_i'$, and
we use:
$$
{\rm Tr}(Y^2) = 5/6 \ ,
$$
$$
{\rm Tr}(\lambda_i \lambda_j) = +1/2\ ,\quad {\rm if} \ i=j \ ,
$$
$$
{\rm Tr}(\lambda_i \lambda_j) = -1/2\ ,\quad {\rm if} \ i\ne j \ ,
$$
$$
{\rm Tr}(T_a T_b) = 2/3 \ , \quad {\rm if}\  a=b \ ,
$$
$$
{\rm Tr}(T_a T_b) = -1/3\ , \quad {\rm if}\  a\ne b \ ,
$$
to obtain
\begin{equation}
24\alpha V(0) = 5\mu_0+ 10\mu_8 \ , 5\mu_0+ 4\mu_8 \ , 
5\mu_0- 8\mu_8 \ , 5\mu_0 -2 \mu_8 \ ,
\label{v011}
\end{equation}
where, the four cases correspond to $(i=j,a=b)$,
$(i=j,a\ne b)$, $(i\ne j,a\ne b)$ and $(i\ne j,a=b)$.
This means that there are two cases which give stable charge
two monopoles for sufficiently small $\mu_0$ and these correspond to
\begin{equation}
|a,+,1>|c,-,1> = |- d,0,2> \ , \quad a \ne c \ ,
\label{chargetwo1}
\end{equation}
and,
\begin{equation}
|c,+,1>|c,-,1> = |cc,0,2> \ .
\label{chargetwo2}
\end{equation}
In the first case, we have used (\ref{bgr}) to write
$a+c$ as $-d$ where $d$ is one of $b,g,r$. We have also
used (\ref{pm}) to set the $SU(2)$ charge to zero.
As examples of these monopoles, we could have a monopole
in the state $|-b,0,2>$ and another in the state
$|bb,0,2>$. The value of $V(0)$ obtained from (\ref{v011})
gives the binding energy of these winding two monopoles.

The charge on the $n=2$ monopole $|- b,0,2>$ is:
$$
M_2  = -T_b + 2Y = {\rm diag} (1,1,0,-1,-1) \ ,
$$
while that on $|bb,0,2>$ is:
$$
M_2 '  = 2T_b + 2Y = {\rm diag} (0,0,2,-1,-1) \ .
$$

The charge three monopoles can be obtained by combining
a charge two and a fundamental monopole. The stability
requirement forces the color combinations to
be $bgr$ and hence there are only two distinct charge
three monopoles:
\begin{equation}
|0,s,3> \ , \quad s=\pm \ .
\label{chargethree}
\end{equation}
In this connection, note that the monopole $|aa,0,2>$
cannot be combined with any fundamental monopole to
yield a stable monopole. For example, if we tried
to combine $|bb,0,2>$ with $|r,+,1>$ to get
$|bbr,+,3>$, this would be unstable to decay into
$|br,0,2>=|-g,0,2>$ and $|b,+,1>$. 

The charge on the $n=3$ monopole $|0,+,3>$ is:
$$
M_3  = \lambda_+ + 3Y = {\rm diag} (1,1,1,-1,-2) \ .
$$

The charge four monopoles can be obtained by combining
two charge two monopoles and this leads to the following
three stable monopoles:
\begin{equation}
|c,0,4> \ , \quad c=b,g,r \ .
\label{chargefour}
\end{equation}
It can be checked that this charge four monopole is stable
to decay into a charge one and a charge three monopole.
Also note that the monopole $|aa,0,2>$ does not combine
in any way with the other monopoles to give another
distinct charge four monopole which is stable. For example,
the combination $|bb,0,2>|gg,0,2>$ is unstable to decay
into $|bg,0,2>+|bg,0,2>$ and the combination
$|bb,0,2>|gr,0,2>$ is equivalent to $|b,0,4>$ which is
included in (\ref{chargefour}).

The charge on the $n=4$ monopole $|b,0,4>$ is:
$$
M_4 = T_b + 4Y = {\rm diag} (1,1,2,-2,-2) \ .
$$

Gardner and Harvey \cite{gh} showed that a charge
five monopole is always unstable to fragmentation
into a charge two plus a charge three monopole.
This follows at once since the charge two and charge
three monopoles only interact via the hypercharge
($Y$) sector in (\ref{intenergy}) and this is always positive.

This then leads us to consider charge six monopoles
and there is only one of them and we can obtain it
by combining two charge three monopoles:
\begin{equation}
|0, 0 , 6>\ .
\label{chargesix}
\end{equation}
Once again the monopole $|aa,0,2>$ cannot be combined
with a charge four monopole to give a stable charge
six monopole. For example, $|bb,0,2>|r,0,4>$ is
unstable to decay into $|br,0,2>=|-g,0,2>$ and
$|b,0,4>$.

The charge on the $n=6$ monopole $|0,0,6>$ is:
$$
M_6 = 6Y = {\rm diag} (2,2,2,-3,-3) \ .
$$

One can easily show that all the monopoles with charge greater
than six are unstable to fragmenting into a monopole of
charge six and something else. This is because the only
interaction of the charge six monopole is via the hypercharge
sector and this contributes positively to (\ref{intenergy}).

This completes the classical stability analysis of
combinations of $SU(5)$ monopoles. Our analysis has uncovered
the spectrum of monopoles that have a chance of being
quantum mechanically stable. That is, so far we have rejected
all monopoles that are classically unstable, or, are classically
stable but still unstable to quantum tunneling. The charge
spectrum and degeneracy of such $H_{SM}$ monopoles
is displayed in Table II where we also tabulate the
spectrum of known fermions and their degeneracy \cite{tv}.
The two spectra show remarkable agreement and so we shall
name each of the monopoles by the symbol for the corresponding
standard model fermion but with a tilde. (For example,
the monopole $|b,+,1>$ will be denoted by ${\tilde u}^b _L$.)
In addition to the monopoles in one-one correspondence
with the standard model fermions, we have also uncovered three
extra monopole states $|cc,0,2>$, $c=b,g,r$. These states are
doubly charged under $SU(3)$ and $U(1)'$ and are $SU(2)$ singlets
and so we shall call them ``diquark monopoles'' and denote them by
${\tilde x}^c$. No correspondingly charged fermions are known to
occur in Nature; as we shall see in Sec. V, these monopoles
will become unstable (metastable) once $SU(3)$ breaks down and 
the colored monopoles get confined. Also, if we assume that
transitions within the same topological sector occur rapidly,
${\tilde x}^c$ will quickly decay into ${\bar d}_R$ since
both are $n=2$ monopoles and ${\tilde x}^c$ is more massive
than ${\bar d}_R$.

Note that the monopoles transform under
$SU(3)$, $SU(2)$ and $U(1)'$ gauge transformations.
By simply counting the degeneracy
of the monopoles, it is clear that all the monopoles
transform in the fundamental representations of the symmetry
groups. This also applies to the diquarks and they should
transform in the fundamental representation of $SU(3)$.


\vfill
\eject

\begin{center}
TABLE II. 
\small 
Charges on classically stable $SU(5)$ monopoles and
on standard model fermions. Also shown are the monopole
degeneracies $d_m$ and the number of fermions with a
given set of charges, $d_f$. The monopoles in the last
row do not have any corresponding fermions in the
standard model.
\end{center} 
\normalsize
\begin{center}
\begin{tabular*}{16.0cm}{|c@{\extracolsep{\fill}}cccc|ccccc|}
\hline
&&&&&&&&&\\[-0.15cm]
                  {$n_{~}$}
                 & {$n_3$}
                 & {$n_2$}
                 & {$n_1$}
                 & {$d_m$}
                 & {$$}
                 & {$SU(3)_c$}
                 & {$SU(2)_L$}
                 & {$U(1)_Y$}
                 & {$d_f$~}             \\[0.20cm]
\tableline 
&&&&&&&&&\\[-0.1cm]
+1&1/3&1/2&+1/6&6&$(u,d)_L   $ &1/3 &1/2  &+1/6&6        \\[0.5cm]
-2&1/3&0  &-1/3&3&$d_R       $ &1/3 &0    &-1/3&3        \\[0.5cm]
-3&0  &1/2&-1/2&2&$(\nu ,e)_L$ &0   &1/2  &-1/2&2        \\[0.5cm]
+4&1/3&0  &+2/3&3&$u_R       $ &1/3 &0    &+2/3&3        \\[0.5cm]
-6&0  &0  &-1  &1&$e_R       $ &0   &0    &-1  &1        \\[0.25cm]
\tableline
&&&&&&&&&\\[-0.15cm]
+2&2/3&0  &+1/3&3 &?            &    &     &    &         \\[0.20cm]
\hline
\end{tabular*}
\end{center}
\vspace{0.3cm}

In the weak coupling case, the monopoles are very
heavy compared to the other mass scales in the problem
($\mu_0$ and $\mu_8$) and the classical result for the
interaction energy given in (\ref{intenergyatzero}) can be
expected to be accurate. Table III shows the 
classical masses of the various monopole bound 
states.

\vspace{0.3cm}
\begin{center}
TABLE III. 
\small
Masses of stable $SU(5)$ monopoles in the weak 
coupling regime. 
\end{center}
\normalsize
\begin{center}
\begin{tabular*}{6cm}{|@{\extracolsep{\fill}}@{  }c@{  }|c|}
\hline
&\\[-0.15cm]
                     {$n_{~}$} & 
                     {Mass} \\[0.20cm]
\tableline
&\\[-0.1cm]
+1&$M_1$                                          \\[0.5cm]
-2&$2M_1+(5\mu_0 /24 - \mu_8 /3)/\alpha$        \\[0.5cm]
-3&$3M_1+(5\mu_0 /8 - \mu_8 /2)/\alpha$        \\[0.5cm]
+4&$4M_1+(5\mu_0 /4 - 3\mu_8 /4)/\alpha$        \\[0.5cm]
-6&$6M_1+(25\mu_0 /8 - 5\mu_8 /4)/\alpha$        \\[0.25cm]
\hline
&\\[-0.15cm]
+2&$2M_1+(5\mu_0 /24 - \mu_8 /12)/\alpha$        \\[0.20cm]
\hline
\end{tabular*}
\end{center}

\vspace{0.3cm}

As the couplings get somewhat larger (while still remaining in the 
weakly coupled regime), we can expect departures
from the classical result for the interaction energy. The
quantum corrections to the interaction energy can (in principle) be 
found by solving the Schrodinger
equation with a reduced mass for the interacting monopoles
and the potential given in (\ref{intenergy}) together with
quantum corrections. Our investigation of this issue is
necessarily incomplete because of the difficulties encountered
in quantizing monopoles. We now describe our calculations and 
partial results.

\

\section{Stable Monopoles: Description of Quantum Effects}

We consider the problem of determining the quantum corrections
to the classical values of the mass of the monopoles shown
in Table III. The full problem involves the quantization of
monopoles and is beyond our present reach. However, the 
problem of determining the quantum bound state energy for 
the potential $V(r)$ in (\ref{intenergy}) can still be
treated analytically within certain approximations.

For monopole masses that are large compared to $\mu_0$ and 
$\mu_3 = 2\mu_8$, the classical results (shown in Table III) should 
be quite accurate. The quantum correction to the classical masses 
involve three factors: (i) the mass of the fundamental monopole 
changes due to quantum corrections, 
(ii) the interaction potential gets quantum corrections, and, 
(iii) the quantum bound state
has higher energy than the classical bound state.
For example, the mass of the $n=2$ monopole will be:
$$
M_2 = 2M_1 + V(0) + \hbar \Delta ( V(r), M_1)
$$
where,
$$
M_1 = M_{1,c} + \hbar \delta M_1
$$
is the mass of the fundamental monopole with first order
quantum corrections,
$$
V(0) = V_0 (0) + \hbar \delta V(0)
$$
is the classical interaction potential at the origin,
$V_0 (0)$, plus its quantum
correction, and, $ \hbar \Delta ( V(r), M_1) $ is the quantum
energy of the bound state of two particles each of mass
$M_1$ and interacting via the potential $V(r)$.
To first order in $\hbar$, we can replace $\Delta (V(r),M_1)$
by $\Delta (V_0(r), M_{1,c})$. Then,
$$
M_2 = 2M_{1,c} +  V_0(r=0) + \hbar \Delta ( V_0(r), M_{1,c})
       + \hbar 2 \delta M_1 + \hbar \delta V(r=0) \ .
$$
In this equation for the $n=2$ monopole mass, we know the
first two terms and will find the third. As far as we
know, magnetic monopoles have not been explicitly quantized
and so the quantum correction to the mass of the fundamental 
monopole, $\hbar \delta M_1$, is
not known. We shall not attempt to quantize the fundamental
monopole here and neither shall we attempt to find the
quantum corrections to the interaction potential 
$\hbar \delta V(r=0)$ (qualitatively discussed below). 

We now turn to evaluating $\Delta ( V_0(r), M_{1,c})$. This can
be obtained by solving the Schrodinger equation
\begin{equation}
-{1\over {2\mu_r^2}} {\vec \nabla}^2 \Psi + V(r) \Psi
= E \Psi \ 
\label{schrodinger}
\end{equation}
where $\mu_r$ is the reduced mass of the two interacting
monopoles and the potential $V(r)$ is given in (\ref{intenergy}).
This Schrodinger equation can easily be solved by numerical methods.
Here, however, we describe an approximation that leads to an
analytic solution.

At weak coupling ($\alpha \rightarrow 0$) we know the 
mass of the fundamental monopole to be
$$
M_{1,c} = {{M_X} \over \alpha} =
{{4\pi} \over g}\sqrt{25\over 2} v_1 \ .
$$
And
$$
{{M_{1,c}} \over {\mu_8}} = {{M_X}\over {\alpha \mu_8}} =
\sqrt{{5} \over {8b}} {{4\pi} \over g} 
$$
where, $g$ is the $SU(5)$ gauge coupling constant. In the
weak coupling regime, $M_1$ is very much larger than
$\mu_3$ and $\mu_0$ and the bound state wave-function can
be thought of as being concentrated at $r=0$ where $V(r)$ has
a global minimum. At somewhat
larger coupling, the bound state wave-function will be
a little more spread out and will experience the interaction
potential at larger $r$. If the distance scale of the 
spread of the wave-function is still small compared to the
other length scales $\mu_3 ^{-1}$ and $\mu_0 ^{-1}$, the
bound state wave-function can be found by expanding the
interaction potential $V(r)$ to linear order in $r$.
Then the Schrodinger equation reduces to:
\begin{equation}
-{1\over {2\mu_r^2}} {\vec \nabla}^2 \Psi + 
(a_0 + a_1 r) \Psi
= E  \Psi \ .
\label{effschrodinger}
\end{equation}
where we have written,
$$
V(r) = a_0 + a_1 r \ .
$$
Here $a_0 = O(\mu_8 / \alpha )$ and 
$a_1 = O(\mu_8 ^2 /\alpha )$. Adopting spherical symmetry, 
the solutions to (\ref{effschrodinger}) are known to be
Airy functions and up to a normalization factor can be
written as:
\begin{equation}
\Psi (r) = {{{\rm Ai}(R-e)} \over r}
\label{soln}
\end{equation}
where 
$$
R = (2 \mu_r a_1)^{1/3} r
$$
and
$$
e = (2 \mu_r a_1)^{1/3} {{(E -a_0 )} \over {a_1}} \ .
$$
In solving the Schrodinger equation, 
we have imposed the boundary conditions that the
wave-function is finite at $r=0$ and vanishes at infinity.
The lowest energy eigenvalue, $E_0$,  is given by the smallest
(in absolute magnitude) root of the Airy function and we get:
\begin{equation}
E_0 = a_0 + 1.86 \left ( {{a_1^2} \over {\mu_r}} \right ) ^{1/3}
\label{eigenvalue}
\end{equation}
Hence:
\begin{equation}
\Delta ( V_0 , M_{1,c} ) = 
           1.86 \left ( {{a_1^2} \over {\mu_r}} \right ) ^{1/3}.
\label{delta}
\end{equation}
with $\mu_r = M_{1,c} /2$ and $a_0$, $a_1$ can be obtained by 
Taylor expanding the potential in (\ref{intenergy}).
Using the dimensional estimate $a_1 = O(\mu_8^2 /\alpha )$
and $M_{1,c} = M_X /\alpha$ gives us
$$
\Delta \sim \mu_8 
       \left ( {{\mu_8} \over {\alpha M_x}} \right )^{1/3} \ .
$$

One can apply (\ref{eigenvalue}) to higher winding monopoles
and find the corresponding $\Delta$. In fact, it is not
at all difficult to numerically solve the Schrodinger equation 
for the full potential $V_0(r)$ and obtain the energy
eigenvalues. But until we can determine the other quantum
corrections, namely $\delta M_1$ and $\delta V(r=0)$, this
exercise cannot yield the desired quantum corrections to the 
monopole mass spectrum.

An estimate of the binding energy of higher winding 
monopoles does not need $\delta M$ but still needs 
$\delta V (r=0)$. To find $\delta V(r=0)$, one might hope to 
draw inspiration from QED where vacuum polarization
effects lead to a modification of the Coulomb interaction
potential between two charges. Here the analogous process
would involve Feynman diagrams with monopole loops. The
calculation of such processes is beyond our reach though these
can probably be ignored at weak coupling. One might also consider 
the effects of loops in the gauge and scalar particle propagators 
due to exchange of electrically charged particles.  The 
contributions of such processes are higher order in coupling
constants and, in any case, cannot shield the monopole
charge since the charge is magnetic. But the processes might 
still yield a magnetic dipole interaction. We hope to consider 
this problem separately.

\section{$SU(3)$ Breaking and Monopole Confinement}

Consider the case when the symmetry of our model is broken down
further when three $SU(3)$ adjoint fields acquire non-parallel
VEVs:
\begin{equation}
[SU(3)\times SU(2)\times U(1)']/Z_6 \longrightarrow
[SU(2)\times U(1)']/Z_2 \ .
\label{su3breaking}
\end{equation}
As the $[SU(2)\times U(1)']/Z_2$ factor is unaffected in this
process, the symmetry breaking is effectively:
\begin{equation}
SU(3) \longrightarrow Z_3 \ .
\label{su3toz3}
\end{equation}
But now,
\begin{equation}
\pi_1 \left ( {{SU(3)} \over {Z_3}} \right ) = Z_3
\label{pi1su3z3}
\end{equation}
and so the symmetry breaking yields $Z_3$ strings. These
strings can terminate on monopoles since the full symmetry
breaking
$$
SU(5) \longrightarrow [SU(2)\times U(1)']/Z_2
$$
does not yield any topological strings.

The monopoles ${\tilde u}_L$ and ${\tilde d}_L$ have 
$SU(3)$ charge equal to $1/3$. Indeed, from Table II,
all the quark-monopoles have $1/3$ $SU(3)$ charge and so each
of the quark-monopoles will get connected to precisely one
string. A string emanating from a monopole can then terminate
on an anti-monopole or split into two strings (since they are
$Z_3$ strings), with each of
the two strings terminating on monopoles. An isolated system
of monopoles and strings must necessarily be an $SU(3)$ singlet.
A picture of a few possible monopole configurations is shown
in Fig. \ref{fig:conf} and these look exactly like the picture of mesons 
and hadrons emerging from the standard model.

The diquark monopoles have $SU(3)$ charge equal to 2/3 and
hence will get connected by two strings.

\section{$SU(3)$ breaking and the Monopole Spectrum}

Once the colored monopoles get connected by $Z_3$ strings,
the stability analysis of Sec. II does not apply. Instead
the long range potential due to the exchange of colored gauge 
and scalar
particles gets replaced by a linear string potential. Hence
\begin{equation}
{\bar V} (r) = {1 \over {4\alpha r}} [
       n_1 n_1' {\rm Tr}(Y^2) (1-e^{-\mu_0 r}) +
     n_3 n_3' {\rm Tr}(\lambda_i \lambda_j) (1-e^{-\mu_3 r})] 
  - n_8 n_8' {\rm Tr}(T_a T_b) \alpha_s  r  \
\label{vrsu3}
\end{equation}
where, $\alpha_s > 0$ is a constant denoting the string tension.
(Interacting monopoles having different $SU(3)$ charges are 
assumed to be connected to each other 
by strings while interacting monopoles with the same $SU(3)$ 
charges are taken to have strings pointing in opposite directions.)
This expression for the potential is only valid on scales
larger than the string thickness; on smaller scales,
the expression in (\ref{intenergy}) should be valid.

As long as a monopole is constructed as a bound state of
two differently colored monopoles, the confinement of colored
monopoles does not affect the stability arguments of Sec. II.
This means that the spectrum contains all the monopoles
corresponding to the standard model fermions.
But things are different for the diquark monopoles since
these are constructed from two identically colored fundamental
monopoles. Hence a diquark monopole must get connected by two 
strings and the long range interaction energy has the form:
\begin{equation}
V_{di}(r) = {1 \over {4\alpha r}} [
 {5\over 6} (1-e^{-\mu_0 r}) - {1\over 2} (1-e^{-\mu_3 r})]
       -{2\over 3} \alpha_s r \ .
\label{su3intenergy}
\end{equation}
Therefore $V_{di} (r)$ is arbitrarily negative as
$r$ becomes large and the diquark monopole is unstable 
to fragmenting into its fundamental monopole constituents.
(The $Z_3$ strings pull the diquark monopole apart into
its quark-monopole constituents.) At short scales, however,
we still expect that the interaction energy (\ref{intenergy})
is applicable and so there is a local minimum of the interaction
energy at $r=0$. This means that the diquark is not 
unstable classically;  but it is metastable
and can decay by quantum mechanical tunneling. This is in
addition to the possible decay of ${\tilde x}^c$ into 
${\bar d}_R$ as mentioned towards the end of Sec. II.

Hence we see that the symmetry breaking (\ref{su3breaking}) 
can serve two purposes: the first is that it can confine the
colored monopoles and the second is that it can make the diquark 
monopole unstable (metastable) and in this way restore the 
correspondence between the $SU(5)$ monopoles and the standard model 
fermions. 

\

\section{$SU(2)\times U(1)$ symmetry breaking by a 
Higgs mechanism}             

The results in Table III bear no resemblance to the mass spectrum 
of the standard model fermions. But there is no reason that we 
should expect any resemblance either. After all, the $SU(2)\times U(1)'$ 
is still unbroken - leading to identical masses for 
${\tilde u}_L$ and ${\tilde d}_L$, for example 
- while, in the standard model, the masses only arise 
from the electroweak symmetry breaking which also lifts the 
degeneracy between members of an $SU(2)$ doublet. 

One way to lift the mass degeneracy between monopoles belonging to an
$SU(2)$ doublet is to break the $SU(2)\times U(1)'$ symmetry further 
to $U(1)_{Q}$ via the Higgs mechanism, i.e. we consider: 
\begin{equation}
G = SU(5) \longrightarrow H_{1} = 
[SU(3) \times SU(2)\times U(1)']/Z_{6} 
\longrightarrow H_{2} = [SU(3) \times U(1)_Q]/Z_3 
\label{ew} 
\end{equation}
This symmetry breaking can be achieved if a field, $\chi$, 
transforming in the fundamental representation of $SU(5)$ gets
a VEV.  The effect of the second symmetry breaking on the $SU(5)$
magnetic monopoles depends on which $U(1)$ subgroup of $H_{1}$ remains 
unbroken. The monopoles with dual electroweak
charge $M$ proportional to $Q$ and quantized via the Dirac
condition with respect to $Q$
are ``aligned'' with the vacuum and remain 
unaffected by the second symmetry breaking. However, the monopoles 
for which the $SU(2)\times U(1)'$ charge is not proportional to $Q$
are ``misaligned'' with the vacuum and cannot remain unaffected by 
the second symmetry breaking since their long range magnetic fields 
are not allowed in the $H_2$ vacuum. For example, 
let us look at the case that the residual $U(1)_Q$ is generated 
by (see eqns. (\ref{lplus}) and (\ref{y})) 
$$
Q= \lambda_- +Y  = 
{\rm diag} \left ( +{1\over 3}, +{1\over 3},+{1\over 3}, -1 ,0 
\right ) \ .
$$ 
The fundamental monopole states $|c,-,1>$, $c=b,g,r$ whose 
magnetic charges are $M = T_{c} + \lambda_{-} + Y $ are clearly 
aligned with $Q$ while the monopoles $|c,+,1>$, $c=b,g,r$ with 
charges $M = T_{c} + \lambda_{+} + Y $ are misaligned. The different
fates of the aligned and misaligned monopoles during the second
symmetry breaking causes the degeneracy between them to 
be lifted\footnote{$SU(5)$ monopoles of winding greater than one 
are found to be always misaligned with the vacuum because the ratio 
of the dual hypercharge and $SU(2)$ charge that
these monopoles carry differs from the ratio occuring in
the generator $Q$ of $U(1)_Q$.}. 

The problem of determining the fate of misaligned monopoles has been 
considered by several authors in the past \cite{bais,shafi,gh}. 
Two outcomes seem possible: (i) the misaligned monopoles get connected 
by strings, or, (ii) they rotate in internal space and get aligned.
Both possibilities are difficult to reconcile with our
general knowledge of topological defects. The first
possibility is in doubt because the symmetry breaking
(\ref{ew}) is known not to yield topological strings. And
there is no known mechanism by which the second possibility
can take place. A third possibility which one of us had envisioned
before \cite{tv} was that the part of the magnetic field of the misaligned 
monopole that is not allowed in the $H_2$ phase gets screened just
as the electric field of a charged particle gets screened in a Higgs phase.
While this remains a possibility when quantum effects are taken into 
account, we will show that it is not allowed at the classical level.

The classic case of misalignment is that of the neutrino-monopole,
${\tilde \nu}_L$. We know that the neutrino has no long range electric
field and hence, if the correspondence between monopoles and standard
model fermions is to be true, we must have a monopole that has no long
range magnetic field. At the classical level, this cannot be true as
we now show in the subsection below. In the subsequent subsection,
we construct an asymptotic solution of a misaligned monopole
connected by a string.

\subsection{Non-existence of a Classical ``Neutrino'' Monopole}

Consider a Yang-Mills theory with a compact semi-simple group $G$ 
undergoing two sequential symmetry breakings:
$$
G \stackrel{\Phi} {\longrightarrow} H_{1}
\stackrel{\chi}{\longrightarrow} H_{2}
$$
If $\Phi$ is in the adjoint representation, $H_{1}$ has the local 
structure of $U(1) \times K$ with $U(1)$ generated by $Q \propto \Phi$ 
and  $K$ some abelian or non-abelian group generated by $K^{\alpha}$, 
$\alpha = 1,2, \cdots Dim(K)$. Although it is always possible to find 
$K^{\alpha}$ which locally satisfy $[K^{\alpha}, \Phi]  = 0$ on the 
asymptotic two sphere,  it is not generally possible to construct 
generators that are {\it globally} well-defined \cite{bala}.
Hence, we will define the generators $K^\alpha$ in coordinate patches
denoted by the letter $P$ where $P=U$ means the ``upper'' patch 
(the whole two sphere except for a small region around the south
pole) and
$P=L$ means the ``lower'' patch (the whole two sphere except for
a small region near the north pole).
Then we can choose $K^\alpha_P$ such that:
\begin{eqnarray}
Tr(\Phi K^{\alpha}_P) = 0\ , \ \alpha = 1,2, \dots Dim(K)\ , \ P=U,L \ .
\label{eq:orth}
\end{eqnarray} 

Let $\Phi^{0}$, $W_{P \mu}^{0}$ be the scalar field and gauge potential
of the monopole solution resulting from the first symmetry breaking. 
Asymptotically, they satisfy:
\begin{eqnarray} 
D_{\mu} \Phi^{0} = 
\partial_{\mu} \Phi^{0} - ig [W_{P \mu}^{0}, \Phi^{0}] = 0 \ .
\label{eq:mono} 
\end{eqnarray}
Now suppose after the second symmetry breaking the scalar field 
$\Phi$ and gauge potential $W_{\mu}$ become:
\begin{eqnarray}
\Phi = \Phi^{0} + \delta \Phi, \; \;\;\;\;
W_{P\mu} = W_{P\mu}^{0} + \delta W_{P\mu} 
\label{eq:change}
\end{eqnarray}
We shall assume that $\Phi$ is in the same topological sector as 
$\Phi^0$ and that it is possible to go to a gauge in which 
$\delta \Phi \propto \Phi^0$. 
In this gauge, it follows from (\ref{eq:mono}) and
\[
D_{\mu} \Phi = \partial_{\mu} \Phi - ig [W_{\mu}, \Phi] = 0
\]
that  $\delta W_P$ can be expressed as:
\[
\delta W_{P\mu} = A_{\mu}\Phi + B_{P\mu}^{\alpha}K^{\alpha}_P \ .
\]
Note that $A_\mu$ is taken to be globally defined since the
$U(1)$ generator $\Phi$ is globally defined.
So now the field strength tensor has the form\cite{corrigan}:
\begin{eqnarray}
G_{\mu\nu} = G_{\mu\nu}^{0} + (\partial_{\mu}A_{\nu} - 
     \partial_{\nu}A_{\mu})\Phi + B_{P\mu\nu}^{\alpha}K^{\alpha}_P 
\label{eq:fs} 
\end{eqnarray}
where $G_{\mu\nu}^{0}$ is the field strength tensor of the monopole
before the second symmetry breaking (constructed out of $W_{\mu}^{0}$)
and the structure of $B_{P\mu\nu}^{\alpha}$ is not important here.
Now let us look at the $U(1)$ magnetic flux (denoted by $h$) 
through a sphere $S_{2}$ at infinity which contains the monopole:
\[
h= \int_{S_{2}} \vec{dS} \cdot Tr(\Phi \vec{B})
\]
where $B_{i} = {1 \over 2} \epsilon_{ijk} G_{jk}$. From (\ref{eq:orth}), 
(\ref{eq:fs}) and Stokes theorem, we have: 
\begin{eqnarray}
h = \int_{S_{2}} \vec{dS} \cdot Tr(\Phi \vec{B})=
     \int_{S_{2}} \vec{dS} \cdot Tr(\Phi \vec{B^{0}}) 
\label{eq:flux}
\end{eqnarray}
where, $ B_{i}^{0} = {1 \over 2} \epsilon_{ijk} G_{jk}^{0}$ and we are
assuming that $A_\mu$ is not the gauge field for a Dirac monopole.

For the monopole to resemble a dualized neutrino after the second
symmetry breaking, we would need $h=0$. But this is not possible
since the right-hand side of (\ref{eq:flux}) is the $U(1)$ magnetic
flux from the original monopole and is non-zero. Therefore a
classical magnetic analog of the neutrino cannot exist.

If we allow ourselves the liberty of including Dirac monopoles,
we could get $h=0$ by cancelling the original monopole flux by
an equal but opposite Dirac monopole flux in the $\Phi$ direction. 
But this configuration with $h=0$ cannot be obtained by a continuous 
evolution of fields starting from the configuration of the original 
monopole. This is essentially because the Dirac monopole satisfies 
the discrete quantization condition and there is no way in which
this discrete condition can be continuously relaxed.

The proof above can be summarized very simply: the monopole from
the first symmetry breaking has a $U(1)$ magnetic flux which
satisfies Maxwell's equations and hence cannot be screened.

\subsection{Monopole Connected by a String}

In this subsection we construct an asymptotic solution
that describes a misaligned monopole connected by a
string. For this purpose, it is cumbersome to consider
the full $SU(5)$ model; instead we consider the simpler
scheme
$$
G = SU(3) \stackrel{\Phi} {\longrightarrow} H = 
[SU(2)\times U(1)']/Z_2  \stackrel{\chi}{\longrightarrow} 
H' = U(1)_Q  
$$ 
where $\Phi$ and $\chi$ are in the adjoint (octet) and 
fundamental (triplet) representations of $SU(3)$.
The form of the self-interaction for $\Phi$ is chosen so that its 
ground state is ``$\lambda_{8}$ like'' ({\it i.e.} related to 
$\lambda_{8}$ by an $SU(3)$ rotation where 
$\lambda_{a}, a=1,2,...8$ are Gell-Mann matrices.) The 
fundamental monopole solution from the first stage has
the configuration  given by
\cite{cofn}:      
\begin{eqnarray}
\Phi^{0} & = & {1 \over 2} \, (\sqrt{3}\psi_{1} - \psi_{2}) 
= {1 \over {2\sqrt{3}}} {\rm diag} \left ( 1,
  {3\over 2} {\vec \sigma} \cdot {\hat r} - {{\bf 1} \over 2}
                                   \right ) \\ 
\label{su3phi}  
W_{\mu}^{0} & = & {i \over e}\,[ \psi_{1}, \partial_{\mu} \psi_{1}]= 
{1 \over 2er}\,
\hat{r} \times \vec{\lambda}'
\label{eq:su3gaugefields} 
\end{eqnarray}
where $e$ is the gauge coupling  and 
\begin{eqnarray}
\psi_{1} = {1 \over 2}\sum_{1}^{3} \lambda_{i}' \hat{r}_{i}, 
 \;\;\;\;
\psi_{2} = {1 \over 2} \lambda_{8}',
\;\;\;\;
\vec{\lambda}' = (\lambda_{1}',\,\lambda_{2}',\,\lambda_{3}')
\label{eq:def} 
\end{eqnarray}
with $\lambda_{i}'$ and $\lambda_{8}'$ defined by:
\begin{equation}
\vec{\lambda}' = {\rm diag} (0, \vec{\sigma}),\,\,\,\,\, \,
\lambda_{8}' = {1 \over \sqrt{3}} {\rm diag} (-2,1,1)
\label{lambdaprime}
\end{equation}
and where $\vec{\sigma}$ are Pauli matrices.
This configuration has the little group $H_{\hat{r}}$ which can be  
obtained from $H_{(0,0,1)} = H$ by conjugation: 
$H_{\hat{r}} = U(\hat{r})\,H \,U^{-1}\!(\hat{r})$, where
$U(\hat{r})$ is the $SU(3)$ transformation which relates
$\Phi ({\hat z})$ and $\Phi(\hat{r})$:
\begin{equation}
U(\hat r ) = 
\left (
\begin{array}{ccc}
1 &0 &0 \\
0 &c &- se^{-i\phi} \\
0 &se^{i\phi} &c 
\end{array}
\right )
\label{ur}
\end{equation}
where $c = \cos{\theta /2}, s = \sin{\theta /2}$ and
$\theta$ and $\phi$ are the usual spherical coordinates.
The generators for infinitesmal transformations of the unbroken
$SU(2)\times U(1)'$ can also be obtained by conjugation:
$$
\tau_{i}(\hat{r}) =
U(\hat{r})\,{1 \over 2} \lambda_{i} \,U^{-1}\!(\hat{r}),\;\; \;
i=1,2,3
$$
$$
\Lambda = U(\hat{r})\, {1 \over 2} \lambda_{8} \,U^{-1}\!(\hat{r})
=  \Phi^{0} \ .
$$ 
(Note that here we are using the Gell-Mann matrices and not the
primed versions defined in (\ref{lambdaprime}).)
As discussed in \cite{bala}, the construction of the off-diagonal
$SU(2)$ generators works for all
but one value of $\hat r$ which we can choose to be on the
negative $z-$axis. However, as we will never need to use the 
$SU(2)$ generators globally, this will not cause us any
problems.

In terms of the diagonal generators of the little group, 
the magnetic charge on the monopole can be expressed as:
\begin{eqnarray}
Q_{M} = - {1 \over e} \psi_{1} = 
{1 \over 2e}(- \sqrt{3} \Lambda + \tau_{3}) \ .
\label{eq:charge}
\end{eqnarray}

Now when the field $\chi$ gets a VEV, the 
symmetry breaks down to $U(1)_Q$ with $Q$ satisfying:
\[ 
Q \chi = 0 \ . 
\]
In the vacuum sector with $\Phi \propto \lambda_8$, 
we will consider two possible VEVs for $\chi$:
\begin{eqnarray}
\chi \propto \left( 
\begin{array}{c}
1 \\
0  \\
0
\end{array}
\right) 
or   
 \left( 
\begin{array}{c}
0 \\
1  \\
0
\end{array}
\right)  \ .
\label{eq:chi0}
\end{eqnarray}
In the monopole sector, the situation becomes more complicated.
For simplicity, 
we only discuss two  cases where, like in the vacuum 
sector, a globally defined charge $Q$ expressible as a
linear combination of the $SU(2)$ and $U(1)$ generators
with constant coefficients
is available. Explicitly,
the VEVs for $\chi$ can be achieved by an $SU(3)$ rotation of the
VEV along the ${\hat z}$ direction:
$$ 
\chi(\hat{r}) = U(\hat{r}) \chi ({\hat z})
$$
where $\chi ({\hat z})$ takes value as in (\ref{eq:chi0}).

At this stage, the asymptotic equations of motion are:
\begin{eqnarray} 
D_{\mu} \Phi  =  \partial_{\mu} \Phi - 
ie [W_{\mu}, \Phi] = 0 
\label{eq:em1} \\
D_{\mu} \chi  =  \partial_{\mu} \chi - 
ie W_{\mu} \chi = 0 
\label{eq:em2} \\
D_\mu G^{\mu \nu}  = 0 
\label{eq:em3}
\end{eqnarray}
where, $G^{\mu \nu}$ is the field strength for the gauge
field corresponding to the unbroken symmetry.
As in the last section, the scalar field $\Phi$ and gauge potential
$W_{\mu}$ can be written as in (\ref{eq:change}) and we can take 
$\Phi = \Phi_{0}$ at infinity.
We now discuss explicit 
solutions to the equations of motion for 
the aligned and misalinged monopoles which correspond to the
first and second choice for $\chi ({\hat z})$ given in
(\ref{eq:chi0}).

\subsubsection{Aligned Monopole}

Let us first consider 
\[                                                
\chi \propto U(\hat{r})
\left( 
\begin{array}{c}
1 \\
0  \\
0
\end{array}
\right)
= \left(
\begin{array}{c}
1 \\
0  \\
0
\end{array}
\right)
\]
and then $Q$ is given by:
\[
Q = {1 \over 2} ( -\sqrt{3} \Lambda + \tau_{3})
\]
which is the same as the charge on the monopole $Q_M$ in
eqn. (\ref{eq:charge}). Hence the charge on the monopole 
is aligned with the vacuum. In this case, the asymptotic equations 
of motion (\ref{eq:em1}), (\ref{eq:em2}) and (\ref{eq:em3})
are satisfied trivially  with $\Phi = \Phi^{0}$ and 
$W_{\mu} = W_{\mu}^{0}$ and the monopole is unaffected by
the second symmetry breaking.

\subsubsection{Misaligned Monopole}

To obtain a misaligned monopole, we would like an asymptotic
configuration for $\chi$ such that the charge $Q$ satisfying
$Q\chi = 0$ is not proportional to the monopole charge
$Q_M$ given in (\ref{eq:charge}). One such $Q$ would be:
\[
Q = {1 \over 2} (\sqrt{3} \Lambda + \tau_{3}) \ .
\]
It is easy to check that this is indeed the charge if
$\chi ({\hat z})$ is chosen to be 
the second possibility in (\ref{eq:chi0}). The configuration
for $\chi ({\hat r})$ can then be found by using the 
$SU(3)$ rotations $U({\hat r})$:
\begin{eqnarray}
\chi \propto U(\hat{r})
\left( 
\begin{array}{c}
0 \\
1 \\
0
\end{array}
\right)
= \left( 
\begin{array}{c}
0 \\
c \\
se^{i\varphi}
\end{array}
\right) 
\label{eq:chi}
\end{eqnarray}
Already it is clear that a string must
get attached to the misaligned monopole since $\chi$
is becoming multi-valued at the south pole ($\theta = \pi$).
We now find the gauge fields corresponding to the $\chi$
configuration in (\ref{eq:chi}).

The misalignment of the monopole means that equation 
(\ref{eq:em2}) is not satisfied trivially with 
$\Phi = \Phi^{0}$ and $W_{\mu} = W_{\mu}^{0}$ as happens 
for the aligned monopole. Using (\ref{eq:change}) 
and (\ref{eq:su3gaugefields}), the $\chi$ equation of motion
(\ref{eq:em2}) now reads  
\begin{eqnarray}
\partial_{\mu} \chi - 
{i \over 2r} \hat{r} \times \vec{\lambda}' \chi - 
ie \delta W_{\mu} \chi = 0 
\label{eq:W}
\end{eqnarray}
with $\chi$ given as in (\ref{eq:chi}).
The solution to eq. (\ref{eq:W}) and 
(\ref{eq:em1}) can be readily found:
\[
\delta W_{\mu} = {1 \over er} \tan{\theta \over 2} 
\ {\hat e}_{\phi}\ \psi_{1} + a_{\mu} \ Q
\]
where $\psi_{1}$ was given in (\ref{eq:def}), ${\hat e}_{\phi}$ 
is the unit vector in the $\varphi$ direction and $a_{\mu}$ 
is an arbitrary four vector function. Notice that the gauge 
potential has developed a Dirac monopole type potential where 
there is a line singularity along the south pole. 

The field strength tensor is now given by 
\begin{eqnarray*}
G_{\mu\nu} & = & \partial_{\mu}W_{\nu} - 
           \partial_{\nu}W_{\mu} - 
          ie [W_{\mu}, W_{\nu}] \\
 & = & 0 + {1 \over e} f_{\mu\nu}Q \\
f_{\mu\nu} & = & \partial_{\mu}a_{\nu} - 
              \partial_{\nu}a_{\mu} \ .
\end{eqnarray*} 
Then the gauge field equations, (\ref{eq:em3}),
reduce to $U(1)_Q$ field equations,
\[
\partial_{\mu}f^{\mu\nu} = 0 
\]
except at the south pole where there is a singularity. 

The fact that the field strength $G_{\mu \nu}$ is 
proportional to $Q$ shows that the only long range
magnetic fields are those allowed by the vacuum. The other
fields must be screened but at the cost of introducing
a string which, in our case, is along the negative
$z-$axis. For example, in the case that the original
monopole had vanishing field strength in the $Q$ direction,
we could take $a_\mu =0$ and the monopole long range
field would be completely screened but for the string
attached to it. In this case, the above solution (in an
asymptotic region excluding the south pole) is
pure gauge with 
$$
W_{\mu} = - {i \over e}\partial_{\mu} 
U(\hat{r})U^{-1}(\hat{r}) \ .
$$

The presence of the string attached to misaligned monopoles 
seems to be forced on us by the presence 
of the $SU(3)$ monopole even though the string is not 
topological. This might lead us to think that the string can 
also terminate on something else besides an $SU(3)$ antimonopole. 
An obvious candidate for the terminus of the string is
Nambu's electroweak monopole \cite{nambu}. 

Imagine the string connected to the $SU(3)$ monopole to be very 
long and along the $-z$ axis. In the region 
far from the $SU(3)$ monopole and in the vicinity of the string 
we have 
$$
\Phi = {1 \over {2\sqrt{3}}} {\rm diag} (1,-2,1) \ .
$$
In this region, the string is like an electroweak $Z-$string
\cite{tanmay} and can terminate on an electroweak magnetic
monopole \cite{nambu}. The $\chi$ configuration of Nambu's
electroweak monpole is purely in the unbroken $SU(2)$ sector 
and is given by:
\begin{equation}
\chi = 
\left(
\begin{array}{c}
{\rm sin} (\theta ' /2) \\
0 \\
{\rm cos}(\theta ' /2) e^{i\varphi}
\end{array}
\right)
\label{nambupole}
\end{equation}
where, $\theta '$ is the spherical azimuthal angle measured
from the $+z$ axis and with Nambu's monopole as
the origin.

It would be satisfying to construct a globally well-defined
configuration that represents an $SU(3)$ monopole connected
by a $Z-$string to an electroweak monopole. However, we have
not succeeded in constructing such a configuration. The
problem is that we need to match the configurations
in (\ref{eq:chi}) and (\ref{nambupole}) in a smooth
manner. However, the two configurations involve different
components of $\chi$ and this makes the matching difficult.
If a smooth matching turns out to be impossible, it would suggest 
that the strings connecting misaligned monopoles should be 
considered to be of topological origin within the low energy 
theory whose symmetry group is $H$ in the one monopole sector. 
In other words, the string connecting the $SU(3)$ monopole 
would be unstable {\it only} to breaking via the formation of an 
$SU(3)$ monopole-antimonopole pair.

\subsection{Resort to Strong Coupling?} 

Our conclusion then suggests that the correspondence between
$SU(5)$ monopoles and standard model fermions fails at weak dual
electroweak coupling within the realm of classical considerations.
This, however, is not fatal to the correspondence because
we know that the electroweak model is weakly coupled and
hence the dual electroweak model must be strongly coupled
(see Table I). So as far as the construction of the dual standard 
model is concerned, we must necessarily include strong coupling
quantum effects in the dual electroweak symmetry breaking 
and the classical analyses described above
may not be indicative of the true picture. 
While an analysis of strong coupling effects in the dual
electroweak sector is beyond the scope of the present
paper, we would like to mention that the considerations
in \cite{banks} seem to be relevant. We hope to return to this 
issue some time in the future.

\section{$\theta$ angle, spin and statistics}

It is well-known that the presence of isospinor bound
states can lead to spin on a monopole \cite{jr} 
with the usual spin-statistics connection \cite{goldhaber}. 
If the model leading to the monopole
includes a $\theta$ term, the condition for determining
whether the monopole (dyon) is a fermion or a boson 
is simply \cite{wilczek} that if
\begin{equation}
\Sigma \equiv {{(m q - 2\theta )} \over {4\pi}} 
\label{fermioncondition}
\end{equation}
is half-integral (integral) then the dyon is a fermion (boson).
($m$ is the magnetic charge and $q$ is the total electric charge
on the dyon.) Note that $q$ contains the $\theta$ contribution
to the electric charge of the dyon \cite{witten}. Since this
contribution is $e\theta/2\pi$, where $e$ is the electric charge 
of the adjoint field, the $\theta$ contributions cancel
in (\ref{fermioncondition}) and so the statistics of the dyon
does not depend on $\theta$.

\begin{figure}[h]
\centerline{\psfig{file=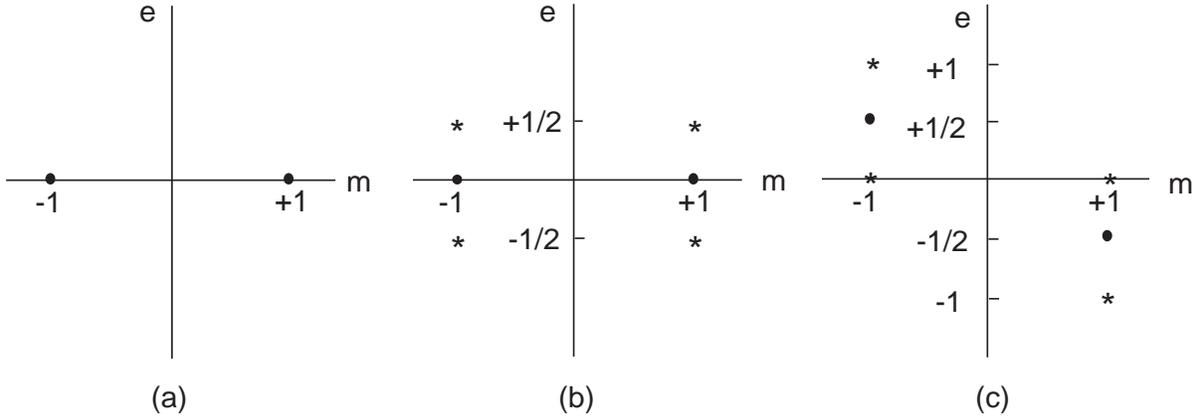,width=16.0cm,angle=90}}
\caption{ The positions of fermionic (represented by $\star$) and bosonic 
(filled circles) 
dyons on the electric ($e$) and magnetic ($m$) charge plane.
Fig. (a) shows the positions for two purely magnetic charges
in the case when there are no electric bound states and $\theta =0$.
Fig. (b) shows the positions of the dyonic states when there
are electric bound states with electric charge $1/2$. 
As shown in Fig. (c), on introducing $\theta = - \pi$, the positions 
of the fermionic and bosonic particles shift so that the purely magnetic 
and, presumably the lightest states, are fermionic. (The full
spectrum of dyons will contain other states as well which we 
have not shown.) }
\label{fig:spec}
\end{figure}

If we include an $SU(5)$ fundamental scalar field $\chi$, 
we would expect that bound states of $\chi$ with the monopoles will 
convert the monopoles to dyons with  spin half integral.
For illustrative purpose, let us look at a simpler example - 
the `t Hooft-Polyakov monopole with an isospinor field $U$ along 
the lines of Ref. \cite{jr}. In the presence of the field
$U$, the two fundamental monopoles (monopole and antimonopole) 
give rise to a tower of dyonic states in which the smallest
electric charge is $1/2$. The smaller charge states are shown in 
Fig. \ref{fig:spec}b and these include the two monopole states with 
zero electric charge and zero spin and four dyonic states with 
electric charge $\pm 1/2$ and spin $1/2$. The four spin-1/2 dyonic 
states can be written as $(\pm 1,\pm 1/2)$ where the first
component is the magnetic charge and the second the electric
charge, and they are degenerate in mass. This situation is 
unsatisfactory since the states with the lowest energy are bosonic,
and even if we identify, say, the spin-$1/2$ $(+1,+1/2)$ dyon with 
$u_L$, the $(-1,-1/2)$ state would be ${\bar u}_L$ but then there would 
be two other dyonic states, $\pm (1,-1/2)$,
with the same mass that are not seen in Nature. The 
quadruple mass degeneracy is actually a consequence of P and CP 
invariance since a parity transform of $(+1,+1/2)$ yields
$(-1,+1/2)$ and a CP transformation of $(+1,+1/2)$ yields 
$(+1,-1/2)$. Hence, it is clear that if we want to break the 
mass degeneracy of these four dyonic states, we must also
break the P and CP connection between them.

A term that gives rise to P and CP violation is a $\theta$ term
($\theta \ne 0 , \pi$). If such a term is included in the action,
the quantization of the charge changes \cite{witten}, and the 
lowest lying states become: 
$$
{\vec q}_{+0} = 
\left ( +1, {\theta \over {2\pi}} \right ) \ ,
$$
$$
{\vec q}_{-0} = 
\left ( -1, -{\theta \over {2\pi}} \right ) \ ,
$$
$$
{\vec q}_{++} = 
\left ( +1, +{1\over 2}+{\theta \over {2\pi}} \right ) \ ,
$$
$$
{\vec q}_{+-} = 
\left ( +1, -{1\over 2}+{\theta \over {2\pi}} \right ) \ ,
$$
$$
{\vec q}_{-+} = 
\left ( -1, +{1\over 2}-{\theta \over {2\pi}} \right ) \ ,
$$
$$
{\vec q}_{--} = 
\left ( -1, -{1\over 2}-{\theta \over {2\pi}} \right ) \ .
$$
Now since the magnetic to electric charge ratios are identical
for ${\vec q}_{++}$ and ${\vec q}_{--}$ but different from
the ratio for ${\vec q}_{+-}$ and ${\vec q}_{+-}$, we only expect
two-fold degeneracy of the masses in fermionic states. 
An especially interesting case is when $\theta =\pi$ (see Fig. 
\ref{fig:spec}c). In this
case, P and CP violation are absent from the model since 
$\theta =\pi$ and $\theta = -\pi$ are related by a $2\pi$ shift.
However, the four-fold degeneracy between the fermionic dyons 
is still broken and  now the lightest dyonic states (of zero electric 
charge) are fermionic and only two-fold degenerate.

Note that in claiming that there is only two-fold degeneracy of
the dyons, we are using the result that the dyons with smaller
electric charge in Fig. \ref{fig:spec} have smaller mass.
This follows from the standard semiclassical quantization 
of monopoles. Suppose we consider a classical monopole configuration 
(including the $U$ field) and we find that
it has a  mass $M_0$ where the subscript means that it
has vanishing electric charge. Upon semi-classical quantization,
to lowest order, the monopole's mass becomes:
\begin{eqnarray}
M_{q} = M_{0} + {q^2 \over {2I}} + \delta M, 
\label{eq:quan}
\end{eqnarray}
where $q^2 / {2I}$ comes from the quantization of the dyonic collective 
coordinate. Here $I > 0$ is a moment of inertia associated with the 
dyonic rotor degree of freedom. (For the quantization of the
monopole see \cite{tw,hr}; for reviews of the quantization procedure, 
see \cite{godolive}.) The $\delta M$ term includes
the zero-point energy from the quantization of small fluctuations 
around the original classical configuration and the possible 
contributions from the quantization of collective coordinates
other than the dyonic rotor collective coordinate. These
contributions only depend on the classical background that
is being quantized and the values of the other quantum numbers - other
than the electric charge - that the dyon might carry. 
So $\delta M$ is the same for all the
dyons that we are considering since these only differ in the
value of their electric charge.
Then from eq. (\ref{eq:quan}) it is clear that the dyons
with smaller electric charge in Fig. \ref{fig:spec}b have smaller mass.

To summarize this section, we can convert fundamental monopoles
into fermions by considering isospin bound states on the 
monopoles that also convert them into dyons. This process leads
to a four-fold mass degeneracy of dyons that can be reduced to
a two-fold degeneracy by including a $\theta$ term in the
action. In the special case when $\theta =\pi$, pure
monopoles (with vanishing electric charge) can be fermions. This 
seems to us like the most attractive way for making the monopoles 
to be fermions as well as to only have two-fold (rather than 
four-fold) degeneracy\footnote{Note also that the stability of
the higher winding monopoles as discussed in Sec. II is unaffected
in this scenario since the electric charge vanishes.}. 
A potential difficulty in this scenario - one that we are 
aware of - is that it may not be possible to get
the higher charge monopoles to have spin 1/2 if the fundamental 
monopoles have spin 1/2 since the combination of two spin 1/2
particles is expected to yield an integral spin object.
In future work we plan to confront this difficulty and to
determine the spectrum of stable dyons in an $SU(5)$ model with 
a $\chi$ field.

\section{Conclusions and Discussion}

We have found the spectrum of stable $SU(5)$ monopoles
and established a correspondence with a single family of fermions 
in the standard model. In addition to the monopoles in one-one 
correspondence with the standard model fermions, we find an additional
monopole which we have termed the ``diquark'' monopole. 
The charges and classical masses of the stable monopoles are
displayed in Tables II and III.

\begin{figure}[h]
\centerline{\psfig{file=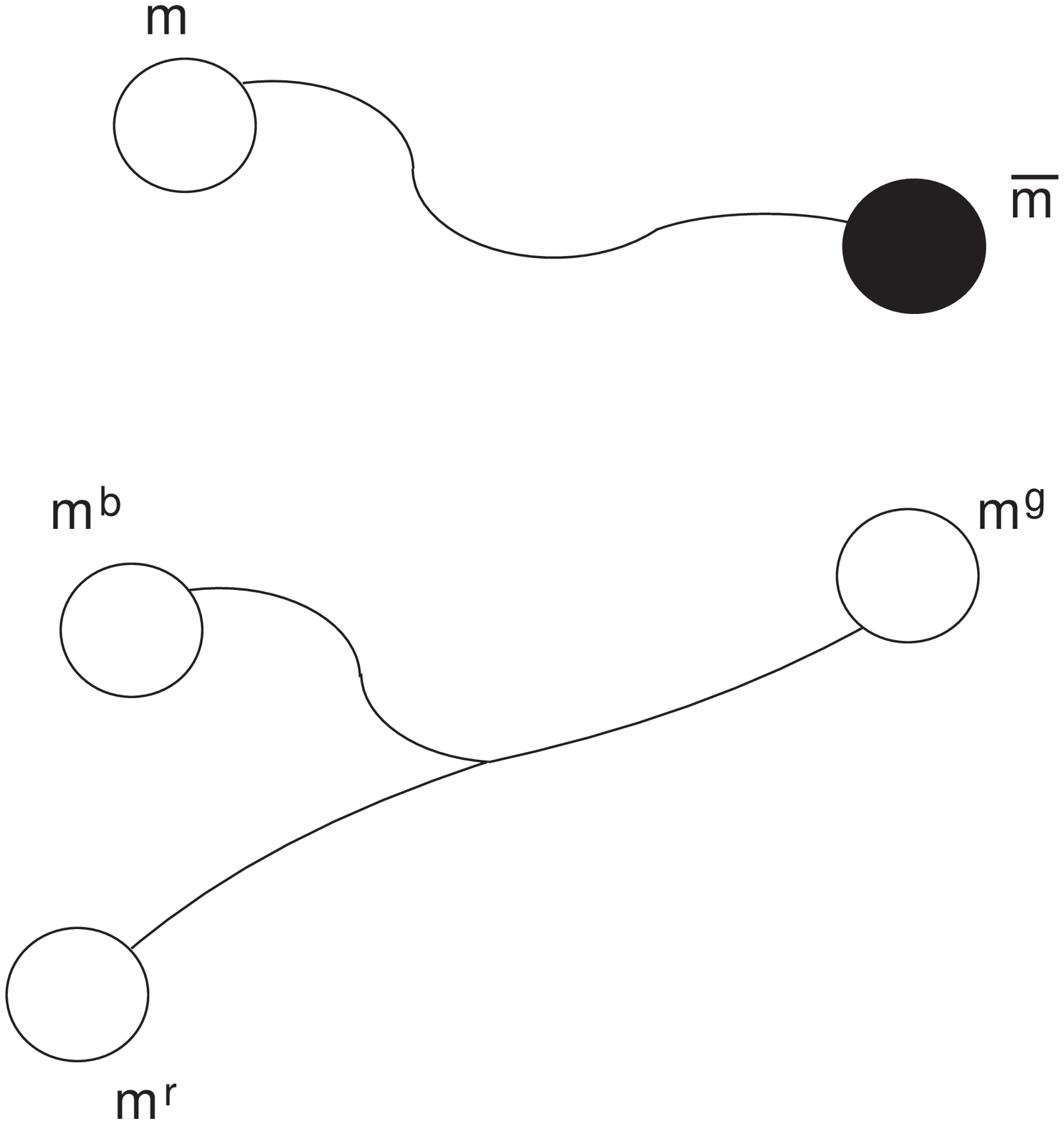,height=6.0cm}}
\caption{Examples of confined clusters of monopoles}
\label{fig:conf}
\end{figure}

Cofinement of the colored monopoles is achieved by breaking the
$SU(3)$ factor of the little group by a Higgs mechanism. This
symmetry breaking also causes the diquark monopole to become
unstable (metastable) and the correspondence
between the charge spectra of monopoles and a single family of
standard model fermions is exact. The picture of confined
monopoles mimics the confinement of quarks as shown in 
Fig. \ref{fig:conf}.

At this stage, the masses of the monopoles occuring in an
$SU(2)$ doublet are degenerate. To lift this degeneracy, it
is necessary to break the electroweak factor 
($[SU(2)\times U(1)']/Z_2$) of the symmetry group.
We have investigated the consequences of achieving the 
symmetry breaking by a Higgs mechanism within the domain
of classical physics. This has the undesirable effect that
certain monopoles which are misaligned with the final
vacuum get connected by strings. This is in spite of the
fact that the electroweak symmetry breaking does not
admit topological strings. Then for the correspondence 
between the monopoles and the standard model
fermions to survive, we need to break the
$SU(2)\times U(1)'$ factor by a mechanism other than
the Higgs mechanism. An alternative may be to still use the
Higgs mechanism to break the symmetry but, in addition, to 
consider a scheme which can screen the ensuing strings. Such 
phenomena are believed to occur in strongly coupled 
theories with some supporting evidence in supersymmetric
theories \cite{intril}. The success of the monopole-fermion 
correspondence depends quite crucially on the determination of 
whether such a scheme exists in our context.

The undesirable consequence of the dual electroweak symmetry
breaking on the misaligned monopoles is actually quite remarkable
in light of particle physics as we know it. We know that at low
energies, the electroweak coupling is weak and the strong ($SU(3)$) 
coupling is strong. This means that a dual model must have the reverse
situation - the dual electroweak coupling must be strong and the
dual strong coupling must be weak (see Table I). Then the
strong coupling problem (quark confinement) that we encounter
in the standard model must have a counterpart problem
in the dual standard model. Furthermore, this problem must arise
in the dual electroweak sector - just as we observe it to arise.
This agreement between the two dual models seems to further support 
the idea that they are different descriptions of the same 
physics ({\it i.e.} two sides of the same coin).

The idea underlying our attempts to establish a monopole
version of the standard model is that it may help us
to understand various features of the standard model.
These would include the charge spectrum of fermions and the
representations in which the fermions occur.
In addition, the model could be a natural description of the real
world at strong coupling. Then the model would be useful
to describe the confinement of quarks since this is a
phenomenon associated with strong coupling in the standard
model and hence weak coupling in the dual model. If the
strong coupling aspects of the electroweak sector are easier
to understand - 
perhaps in a lattice gauge theory context \cite{banks}  - 
than strong coupling in the QCD sector, the
dual standard model may yield a short cut to the simultaneous 
understanding of both the strongly and weakly coupled sectors
in the standard model. 

If the dual picture is correct, we can try and make predictions
based on essentially classical considerations
that may be experimentally testable. In the QCD sector, at
least, the high energy scattering of particles should resemble the 
scattering of monopoles. Assuming that the results presently 
available for $SU(2)$ monopoles \cite{manton} also apply to
monopoles in other theories, quarks in head-on (zero angular 
momentum) collisions should scatter at $90^o$; at larger impact 
parameters ({\it i.e.} larger angular momentum), the collisions 
should lead to products with greater transverse energies than 
predicted by point-like interactions. Another qualitative prediction 
of the $SU(5)$ model would be the existence of a metastable diquark. 

Numerous issues remain to be resolved if we are to be able to think 
of standard model fermions as magnetic monopoles. These
issues include determining the mass spectrum of the monopoles, 
getting the monopoles to have the right spin and statistics,
introducing chirality in the monopoles and obtaining the family 
structure of the standard model \cite{hlgstv}.  At present, all these 
issues (and many more) are open for investigation. 

\

\

\noindent{\bf Acknowledgements}

We wish to thank Jeff Harvey, Lawrence Krauss, John Preskill and 
Glenn Starkman for discussions. TV was supported by the DOE.


\renewcommand\floatpagefraction{.9}
\renewcommand\topfraction{.9}
\renewcommand\bottomfraction{.9}
\renewcommand\textfraction{.1}


\vspace{1.6cm}
\renewcommand\floatpagefraction{.9}
\renewcommand\topfraction{.9}
\renewcommand\bottomfraction{.9}
\renewcommand\textfraction{.1}

\end{document}